\documentstyle[12pt,epsf]{article}
\newcommand{\la}[1]{\label{#1}}
\newcommand{\be}{\begin{equation}}
\newcommand{\ee}{\end{equation}}
\newcommand{\ba}{\begin{eqnarray}}
\newcommand{\ea}{\end{eqnarray}}
\newcommand{\nr}[1]{(\ref{#1})}
\newcommand{\htt}{\hat t}
\newcommand{\hp}{\hat p}
\newcommand{\hq}{\hat q}
\newcommand{\hb}{\hat \beta}
\newcommand{\hqc}{\hat q_c(\htt)}
\newcommand{\hE}{\hat E}

\newcommand{\dn}{{\rm dn}}
\newcommand{\sn}{{\rm sn}}
\newcommand{\cn}{{\rm cn}}
\newcommand{\qmax}{\hat q_{\rm max}}
\newcommand{\ReK}{\mathop{\rm Re}K(k)}
\newcommand{\lsi}{\raise0.3ex\hbox{$<$\kern-0.75em\raise-1.1ex\hbox{$\sim$}}}
\newcommand{\gsi}{\raise0.3ex\hbox{$>$\kern-0.75em\raise-1.1ex\hbox{$\sim$}}}
\newcommand{\lsim}{\mathop{\lsi}}
\newcommand{\gsim}{\mathop{\gsi}}

\makeatletter \@addtoreset{equation}{section} \makeatother
\renewcommand{\theequation}{\arabic{section}.\arabic{equation}}

\begin{document}
 
\setlength{\baselineskip}{0.6cm}
\newcommand{\nn}{\nonumber}
\newcommand{\tr}{{\rm Tr\,}}
\newcommand{\fr}[2]{{\frac{#1}{#2}}}
\newcommand{\figysize}{16.0cm}
\newcommand{\figtopspace}{\vspace*{-1.5cm}}
\newcommand{\figbottomspace}{\vspace*{-5.0cm}}
  
\begin{titlepage}
\begin{flushright}
HD-THEP-97-18\\
hep-ph/9705312\\
May 14, 1997
\end{flushright}
\begin{centering}
\vfill
 
{\bf 
The finite temperature real time $\hbar^2$ corrections
in quantum mechanics}
\vspace{1cm}
 
D. B\"odeker\footnote{bodeker@thphys.uni-heidelberg.de}, 
M. Laine\footnote{m.laine@thphys.uni-heidelberg.de} and
O. Philipsen\footnote{o.philipsen@thphys.uni-heidelberg.de} \\
 
\vspace{1cm} {\em 
Institut f\"ur Theoretische Physik, 
Philosophenweg 16, 
D-69120 Heidelberg, Germany}
 
\vspace{2cm}
 
{\bf Abstract}
 
\vspace{0.5cm}

We study non-perturbative real time correlation functions at finite
temperature.  In order to see whether the classical term gives a 
good approximation in the high temperature limit $T\gg \hbar\omega$,
we consider the first $\hbar^2$ quantum corrections.
We find that for the simplest non-trivial case, the quantum mechanical
anharmonic oscillator, the classical result is reliable
only for moderately large
times: after some time $t_*$ the classical approximation breaks down 
even at high temperatures. Moreover, the result for the first quantum
corrections cannot, in general, be reproduced by modifying the
parameters of the classical theory.
\end{centering}

\vspace{0.5cm}\noindent


\vspace{0.3cm}\noindent
 
\vfill \vfill
\noindent
 
\end{titlepage}
 

\section{Introduction}

Real time processes at finite temperature play an essential role in
the physics of the early universe and of heavy ion collisions. A key
quantity in scenarios of baryogenesis~\cite{kuzmin,rs} is the rate for
electroweak baryon number violation (the sphaleron rate). In the broken
phase the sphaleron rate can be computed with semiclassical
methods~\cite{kuzmin,ar,khlebnikov} but in the symmetric phase
\cite{p} they are not reliable.  Unfortunately, a direct
non-perturbative lattice determination of the hot sphaleron rate is
not available, either. 

The most promising approach to this problem~\cite{grigoriev} is to
compute the sphaleron rate in a classical real time simulation since
the relevant thermal transitions are essentially
classical. Considerable work has been done in this
direction~\cite{ambjorn91}--\cite{smit}.

Treating the dynamics of a classical gauge field system one 
is nevertheless faced
with severe difficulties~\cite{nielsen}--\cite{arnold97}. 
The high momentum modes with
$k\gsim T$ which do not behave classically, do not decouple from the
dynamics. In general, these modes lead to ultraviolet divergences in
the classical correlation functions which cannot be removed by
introducing local counterterms in the classical 
theory~\cite{bodeker,arnold97}. 

There is
another question related to the classical approach which
has hardly been
considered so far: 
under which conditions is the classical
approximation for the low momentum modes 
reliable?  One systematic way
of investigating the validity of the classical approximation is to
compute the first quantum corrections in the $\hbar$-expansion. So
far, the expressions have been derived only for quantum mechanics and
scalar field theories~\cite{b}.  However, these simple cases should
already teach us something in spite of the fact that topological
observables and the associated rate do not exist.  In these models
relevant observables might be related for instance to the damping
rate~\cite{aarts,smit}.

The purpose of the present paper is to evaluate the quantum 
corrections in the simplest non-trivial case, the 
quantum mechanical anharmonic oscillator. This study serves to  
estimate the feasibility of similar studies in field theories. 
Moreover, we believe that some of the general results might
be carried over to that context.

We find that while at small times the classical approximation
is reliable, it breaks down at large enough times. The reason
is that the functional form of the quantum corrections is qualitatively 
different from that of the classical answer, in a way which
cannot be accounted for by modifying the parameters of the
classical result.

The paper is organized as follows.
In Sec.~\ref{Formulation} we discuss the formulation of the problem.
In Sec.~\ref{sec:ho} we briefly discuss the harmonic oscillator 
and in Sec.~\ref{aho} the anharmonic oscillator. The ``symmetric''
and ``broken'' cases of the latter are analyzed
in more detail in Secs.~\ref{symmetric}, \ref{broken}, 
and we conclude in Sec.~\ref{concl}.

\section{The formulation of the problem}
\label{Formulation}

We consider one bosonic degree of freedom $q$ with
conjugate momentum $p$ and the Hamiltonian 
\be
H=\frac{p^2}{2}+U(q), 
\ee
where
\be
U(q)=\left\{\begin{array}{l}
+\frac{1}{2}\omega^2q^2+\frac{1}{4}g^2 q^4 
\\
-\frac{1}{2}\omega^2q^2+\frac{1}{4}g^2 q^4+\frac{\omega^4}{4 g^2} 

\end{array}\right. .
\label{U}
\ee
We refer to the two cases of a positive and of a negative quadratic
term as the symmetric and the broken case, respectively.
Quantum mechanical (Heisenberg) operators are denoted by capital letters,
for example
\be
Q(t)=e^{\frac{i}{\hbar} H t} Q(0) e^{-\frac{i}{\hbar} H t}.
\ee
The finite temperature correlator we consider is 
\be
C(t)=
\left\langle 
\frac{1}{2}\Big[
Q(t)Q(0)+Q(0)Q(t)
\Big]
\right\rangle= \frac{1}{Z} \mathop{\rm Re}\mathop{\rm Tr} 
\left[e^{-\beta H(P,Q)} Q(t)Q(0)\right], \la{c}
\ee
relevant for the time dependence of
\be
\left\langle
\Big[Q(t)-Q(0)\Big]^2
\right\rangle.
\ee
Here $Z= \mathop{\rm Tr}\left[\exp(-\beta H)\right]$ and 
$\beta$ is the inverse temperature.
Note that $C(t)$ is an even function of $t$.

In~\cite{b}, the expansion
\begin{eqnarray}
  C(t) = C_{\rm cl}(t) + C_\hbar(t) +  C_{\hbar^2}(t) + 
  {\cal O}(\hbar^3)
\end{eqnarray}
was derived for $C(t)$. The classical result 
is~\cite{dolan,bochkarev,bodeker}
\begin{eqnarray}
  C_{\rm cl}(t)= Z_{\rm cl}^{-1} \int \frac{dp dq}{2\pi\hbar}
  e^{-\beta H(p,q)}  q q_{\rm c}(t) ,
  \la{c0}
\end{eqnarray}
where $Z_{\rm cl} = \int \frac{dp dq}{2\pi\hbar} e^{-\beta H(p,q)}$
and $q_c(t)$ is the solution of the classical equations of motion
with the initial conditions $q_c(0)=q,$ $ \dot{q}_c(0)=p$. 
This expression corresponds to the prescription suggested 
by Grigoriev and Rubakov~\cite{grigoriev}. 

As for the quantum corrections, 
the contribution $C_\hbar(t)$ vanishes. The result to 
order $\hbar^2$ is then~\cite{b}
\ba
  C(t) &=& Z^{-1} \int \frac{dp dq}{2\pi\hbar}  e^{-\beta H(p,q)} 
  \biggl\{ \biggl[ 1  - \frac{\hbar^2\beta^2}{24}  U''(q) +
  \frac{\hbar^2\beta}{24} \left(
  \partial_q^2 
  +  U''(q) \partial_p^2 
 \right) \biggr] q q_{\rm c}(t) \nn \\
& & \hspace*{2cm}  - \frac{\hbar^2}{24}
q    \int_0^t dt'  U'''\big(q_{\rm c}(t')\big) 
   \{q_{\rm c}(t'),q_{\rm c}(t)\}_3  \biggr\} + {\cal O}(\hbar^3), \la{ch2}
\ea
where $\{,\}$ denotes the Poisson bracket
\ba
  \{f,g\} = \partial_p f \partial_q g - \partial_p g 
  \partial_q f,
\ea
and
\ba
  \{f,g\}_0= g ,\qquad \{f,g\}_{n+1} = \{f,\{f,g\}_n\}.
  \la{poisson_n}
\ea
Similarly, the expression for $Z$ to order $\hbar^2$ is
\be
Z= \int \frac{dp dq}{2\pi\hbar}  e^{-\beta H(p,q)} 
  \left[ 1  - \frac{\hbar^2\beta^2}{24} U''(q) \right]. \la{z}
\ee

There are thus three kinds of terms 
in the $\hbar^2$-correction to $C(t)$,
denoted by $C_{\hbar^2}^{(i)}(t)$, $i=a,b,c$:
\begin{eqnarray}
  C_{\hbar^2}(t) = C_{\hbar^2}^{(a)}(t) + C_{\hbar^2}^{(b)}(t) + 
  C_{\hbar^2}^{(c)}(t),
\end{eqnarray}
where
\begin{eqnarray}
 C_{\hbar^2}^{(a)}(t) & = & Z_{\rm cl}^{-1} 
\biggl(\frac{\hbar^2\beta^2}{24}\biggr)
 \int \frac{dp dq}{2\pi\hbar} 
 e^{-\beta H(p,q)}  U''(q)
  \biggl[   C_{\rm cl}(t) -  q q_{\rm c}(t)  \biggr], \la{cha} \\
  C_{\hbar^2}^{(b)}(t) & = &  Z_{\rm cl}^{-1} 
\biggl(\frac{\hbar^2\beta}{24}\biggr)
 \int \frac{dp dq}{2\pi\hbar} 
 e^{-\beta H(p,q)}
  \biggl[\partial_q^2 
  +  U''(q) \partial_p^2 
 \biggr] q q_{\rm c}(t), \la{chb} \\
   C_{\hbar^2}^{(c)}(t) & = &  Z_{\rm cl}^{-1} 
 \biggl(\frac{- \hbar^2}{24}\biggr)
 \int \frac{dp dq}{2\pi\hbar} 
 e^{-\beta H(p,q)}
   q \int_0^t dt'  U'''\big(q_{\rm c}(t')\big) 
   \{q_{\rm c}(t'),q_{\rm c}(t)\}_3. \la{chc}  
\end{eqnarray}
The term $C_{\hbar^2}^{(a)}(t)$ is a sum of the 
$\hbar^2$ correction to the partition function 
when it combines with the classical result $C_{\rm cl}(t)$,  
and of the corresponding term in the numerator of eq.~\nr{ch2}.
Eqs.~\nr{cha}--\nr{chc} are the corrections we will evaluate below.

One of the key issues of the present
problem is the following: In the
case of static {\it time-independent} correlators, it is 
possible (in a weakly coupled theory)
to reproduce 
the results of the full quantum theory
from a classical theory with a high 
accuracy, provided that the parameters of the 
classical theory are modified appropriately. This is called
dimensional reduction~\cite{ginsparg,kajantie}. 
The question is then 
whether such a resummation might also work 
in the time-dependent case. Indeed, 
it has been proved that the resummation used in the
time-independent context is sufficient 
for making the time-dependent two-point function 
in the scalar $\phi^4$ theory finite 
to two-loop order in perturbation theory
and even for giving 
the corresponding damping rate 
the right leading order numerical 
value~\cite{aarts}. General arguments
in the same direction were also given in~\cite{mt}. The expansion 
in eq.~\nr{ch2} is, in contrast, non-perturbative: 
each term involves contributions from all orders in the coupling constant. 
Let us therefore 
discuss the effects of the resummation in the present 
context (see also~\cite{b}).
Of course, the problem of divergences 
does not occur unlike in field theory.

First, consider dimensional reduction. 
Let us take as an example
the ``symmetric case'' anharmonic oscillator,
\be
U(q)=\frac{1}{2}\omega^2 q^2+\frac{1}{4}g^2q^4.
\ee
The starting point is then a 1-dimensional Euclidean field theory  
defined by 
\be
{\cal L}=\frac{1}{2}(\partial_\tau q)^2+\frac{1}{2}\omega^2q^2+
\frac{1}{4}g^2q^4,\quad
Z=\int {\cal D}q 
\exp(-\frac{1}{\hbar}\int_0^{\beta\hbar}\!d\tau {\cal L}).
\ee
According to dimensional reduction, this can be written as
\be
Z={\rm const.}\times \int \! dq_0 \exp(-S_{\rm eff}),
\ee
where $q_0$ is the zero Matsubara mode. The parameters
in $S_{\rm eff}$ are modified by the non-zero modes. The 
non-zero mode propagator is 
\be
\langle q_n q_m\rangle=
\frac{\delta_{n+m,0}}{\omega^2+(2\pi n T/\hbar)^2}.
\ee
To order $\hbar^2$ (which is a good approximation
as long as $\beta \hbar\omega\ll\pi$), one can then easily
calculate how the mass parameter in the effective theory 
is modified: 
\be
\omega_{\rm eff}^2=\omega^2+3g^2 T\sum_{n\neq 0}
\frac{1}{(2\pi nT/\hbar)^2}=\omega^2+\frac{1}{4}g^2\hbar^2 \beta. \la{effw}
\ee 
The change in the coupling constant is of order $\hbar^4$
and thus does not contribute in the present 
$\hbar^2$-calculation.

Consider, on the other hand,
eqs.~\nr{ch2}, \nr{z}. In eq.~\nr{z}, 
$U''=\omega^2+3 g^2q^2$. The constant $\omega^2$-part
of this expression 
does not contribute in eq.~\nr{ch2} since it is cancelled
by a similar part in the numerator, 
see eq.~\nr{cha}.  
The $q^2$-part, on the other hand, can be exactly 
reproduced by calculating the classical partition 
function $Z_{\rm cl}$ with $\omega^2$ modified according 
to eq.~\nr{effw}: 
\be
\exp\left(-\beta\frac{1}{2}\omega^2q^2\right)
\left(1-\frac{\hbar^2\beta^2}{24}3g^2q^2
\right)=
\exp\left(-\beta\frac{1}{2}
\omega_{\rm eff}^2q^2\right) + {\cal O}(\hbar^4).
\ee
Similarly, the $-\beta^2 U''(q)$-term in the square brackets
in eq.~\nr{ch2} is accounted for by the change in $\omega^2$
according to eq.~\nr{effw}. Thus the term 
$C_{\hbar^2}^{(a)}(t)$ in eq.~\nr{cha} is directly related to
changing the parameters of the classical theory. 
However, there remain the terms 
$C_{\hbar^2}^{(b)}(t)$ and $C_{\hbar^2}^{(c)}(t)$. 
On the other hand, $q_c(t)$ is still a solution to the 
original Hamilton equations of motion. Hence the question is
whether the $\hbar^2$-effects can be  taken into account by determining 
$q_c(t)$ form the equations of motion  with the 
modified parameter $\omega_{\rm eff}^2$
rather than $\omega^2$. 
This issue will be discussed below and we find that, in general, 
such a resummation does not take place. 

Finally, it should be noted that in the field theory
case one is usually interested in a ``rate'' observable: 
a time independent constant determining the time dependence 
of some Green's function, 
for example the sphaleron rate or the damping rate.
We are not aware of such an observable related to $C(t)$
in the present context. We thus consider the 
general large-time functional behaviour of $C(t)$.

\section{Harmonic oscillator}
\la{sec:ho}

In order to show in a simple setting how the $\hbar$-expansion
works and to see what the structure of the perturbative solution
is, let us start by considering briefly the harmonic oscillator.
The classical Hamiltonian is
\be
H=\fr12 p^2+\frac{1}{2}\omega^2 q^2.
\ee
In this trivial case, the correlation function in eq.~\nr{c}
can be calculated exactly, with the result 
\be
C(t)=\frac{\hbar}{2 \omega}
\biggl(\tanh\!\frac{\beta\hbar\omega}{2}\biggr)^{-1} 
\cos \!\omega t. \la{ho}
\ee
Expanding in $\hbar$, one gets 
\be
C_{\rm cl}(t)+
C_{\hbar^2}(t)=\frac{\cos \!\omega t}{\beta\omega^2}
\left[ 1+\frac{1}{12}(\beta\hbar\omega)^2 
\right]. \la{hoh2}
\ee
The fact that it is the symmetric
combination of $Q(t)Q(0)$
which appears in eq.~\nr{c},   
removes the term linear in $\hbar$ from the result.
It is seen that 
the quantum corrections change the amplitude of $C_{\rm cl}(t)$, 
but not the frequency 
since $\omega$ is independent of energy. 
The classical $\hbar^0$-term 
is reliable in the limit $\beta\hbar\omega \ll 1$, 
that is, at high temperatures. At low temperatures, in contrast, 
the $T=0$ result (with $\tanh =1$ in eq.~\nr{ho}) is reliable. 

How is this result reproduced by eqs.~\nr{c0}, \nr{ch2}? 
The solution of the classical equations of motion is 
\be
q_c(t)=q\cos\!\omega t+\frac{p}{\omega}\sin\!\omega t.
\la{hoqct}
\ee
Substituting this into eq.~\nr{c0}, one sees that  
the term proportional to $p$ in $q_c(t)$ does
not contribute due to antisymmetry in $p$, and 
one gets directly
\be
C_{\rm cl}(t)=\frac{1}{\hbar Z_{\rm cl}}
\frac{\cos\!\omega t}{\beta^2\omega^3}
=\frac{\cos\!\omega t}{\beta\omega^2},
\label{CclHO}
\ee
where it was used that $Z_{\rm cl}=(\beta\hbar\omega)^{-1}$.

As for the quantum corrections, 
the last term in eq.~\nr{ch2}
is proportional to the third derivative of the potential 
and thus does not contribute, $C_{\hbar^2}^{(c)}(t)=0$.
The term $C_{\hbar^2}^{(a)}(t)$ in eq.~\nr{cha} does not contribute 
either, since $U''(q)$ is just a constant. There
remains a contribution from $C_{\hbar^2}^{(b)}(t)$ in eq.~\nr{chb},
reproducing eq.~\nr{hoh2}.

\section{Anharmonic oscillator}
\la{aho}

Let us then move to the less trivial case of the anharmonic oscillator.
Here and in the following we use $\omega$, $g$ and
\be
V_0=\frac{\omega^4}{4g^2}
\ee
to introduce the dimensionless variables $\hq,\hp,\htt,\hb,\hE$:
\be
\hq=\frac{g}{\omega}q, \quad
\hp=\frac{g}{\omega^2}p,\quad \htt=\omega t,
\quad \hb = \beta V_0,\quad \hE = \frac{E}{V_0}. 
\la{dimless}
\ee
This rescaling serves to show the parameter
dependence of the final non-perturbative result  more clearly. 
At the same time, it makes the coupling constant equal to
unity so that if one wants to compare with perturbation 
theory, one should go back to the original variables.
In terms of the rescaled variables the potential in eq.~(\ref{U})
reads
\be
U(q)=\left\{\begin{array}{lll}
& V_0 (\hq^4+2\hq^2), & \mbox{\rm the ``symmetric'' case}  \\
& V_0 (\hq^2-1)^2, & \mbox{\rm the ``broken'' case} 
\end{array}\right. .
\ee

A dimensionless combination to which $\hbar$ can 
be attached is
\be
\epsilon = \frac{g^2\hbar}{\omega^3}. 
\ee
The quantity naively
governing the semiclassical expansion is hence $\epsilon^2$.
This is multiplied by some dimensionless function $f(\hb,\htt)$ which
may scale approximately with some power of $\hb$ for given $\htt$.
For instance, in the case of the harmonic oscillator, $f(\hb,\htt)$
scales as $\hb^2$ so that the real expansion parameter is
\be
(\epsilon \hb)^2 \propto (\beta\hbar\omega)^2.
\ee
One of the issues below is how the function $f(\hb,\htt)$ 
behaves in the anharmonic case
as a function of $\hb$. 

With the rescaling performed, one can also write $C(t)$
in a dimensionless form. Factoring out the scale $\omega^2/g^2$,  
the classical correlation function is
\be
C_{\rm cl}(t)=\left(\frac{\omega^2}{g^2}\right)
\left(\frac{\omega^3}{\pi g^2\hbar}\right)\frac{1}{Z_{\rm cl}}
\int_{-\infty}^{\infty}\!d\hp \int_0^{\infty}\!d\hq
e^{-\hb \hE}\hq\hqc \equiv
\left(\frac{\omega^2}{g^2}\right)\hat Z_{\rm cl}^{-1}(\hb) 
\hat C_{\rm cl}(\hb,\htt), \la{cct}
\ee
where
\be
Z_{\rm cl} = \left(\frac{\omega^3}{\pi g^2\hbar}\right)
\hat Z_{\rm cl}(\hb),\quad
\hat Z_{\rm cl}(\hb) = \int_{-\infty}^{\infty}\!d\hp \int_0^{\infty}\!d\hq
e^{-\hb \hE}. \la{cz}
\ee
Here we utilized the symmetry of the integrand in 
$\hq\to -\hq,\hp\to -\hp$. 
We can also write the quantum corrections in a dimensionless form,
\be
C_{\hbar^2}(t)=\epsilon^2
\left(\frac{\omega^2}{g^2}\right)\hat Z_{\rm cl}^{-1}(\hb)
\Bigl[
\hat C_{\hbar^2}^{(a)}(\hb,\htt)+
\hat C_{\hbar^2}^{(b)}(\hb,\htt)+
\hat C_{\hbar^2}^{(c)}(\hb,\htt)
\Bigr]. \la{cq}
\ee
We will then discuss the ``symmetric'' and ``broken'' cases
separately. 

\section{The symmetric case}
\la{symmetric}

\subsection{Numerical results}

The detailed form of the classical solution $q_c(t)$
and of the integrals appearing 
in the symmetric case
is discussed in \ref{appA}.
The expressions to be evaluated are in 
eqs.~\nr{cct}, (\ref{sfzc})--(\ref{sfchc}). We have done 
the evaluation numerically, as well as 
analytically in certain regimes of $\hb,\htt$. 
Let us discuss the numerical result first. The curves are 
displayed in Figs.~\ref{sccl}--\ref{schc}.

\begin{figure}[tb]
 
\vspace*{-1.0cm}
 
\hspace{1cm}
\epsfysize=18cm
\centerline{\epsffile{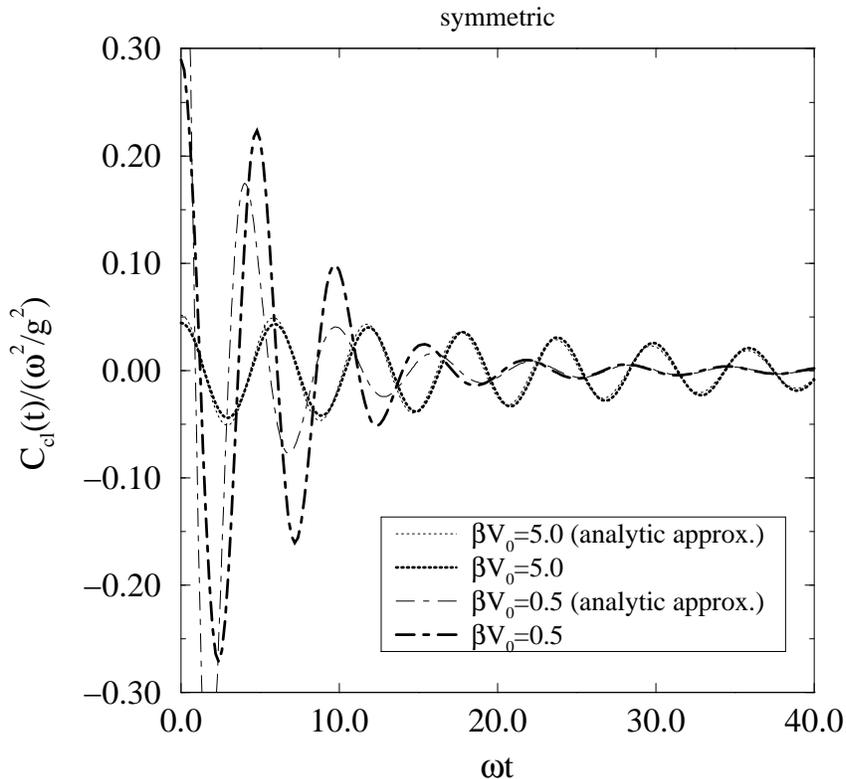}}
 
\vspace*{-6cm}
 
\caption[a]{
The classical correlator $C_{\rm cl}(t)$ in the symmetric case.
The thin line represents the analytic approximation of
Sec.~\ref{ltl}. The difference between the thin and thick lines 
in the regime $1\lsim \omega t\lsim \beta V_0$ is due
to higher order perturbative corrections in $1/\beta V_0 
\sim g^2/\beta\omega^4$.}
\la{sccl}
\end{figure}

\begin{figure}[tb]
 
\vspace*{-1.0cm}
 
\hspace{1cm}
\epsfysize=18cm
\centerline{\epsffile{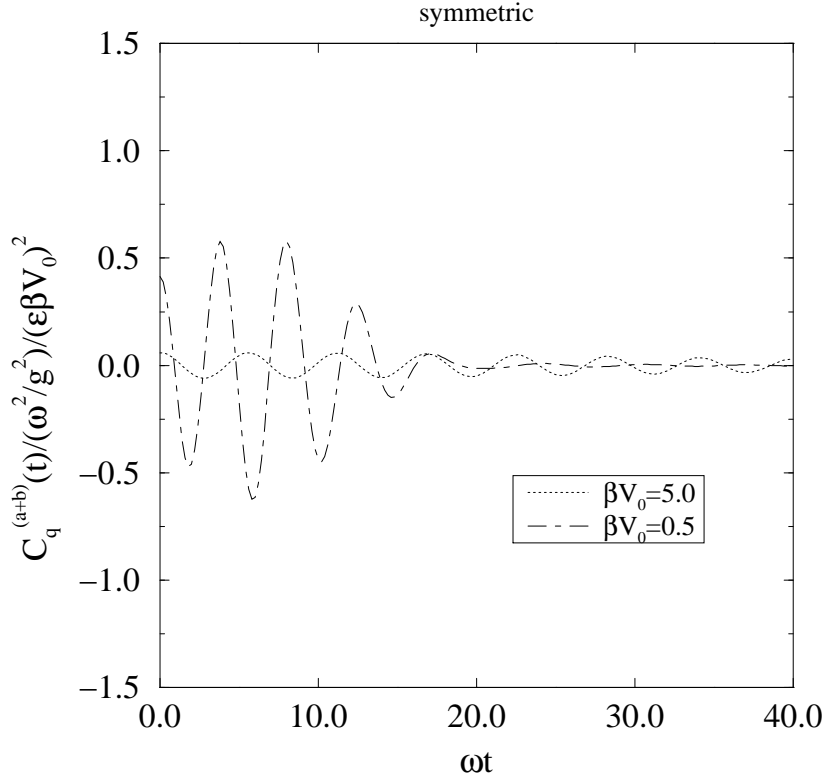}}
 
\vspace*{-6cm}
 
\caption[a]{
The quantum correction 
$C_{\hbar^2}^{(a)}(t) + C_{\hbar^2}^{(b)}(t)$ in the symmetric case.
We have divided out the naive expansion parameter
$(\epsilon\beta V_0)^2=(\fr14 \beta\hbar\omega)^2$.}
\la{ssum}
\end{figure}

\begin{figure}[tb]
 
\vspace*{-1.0cm}
 
\hspace{1cm}
\epsfysize=18cm
\centerline{\epsffile{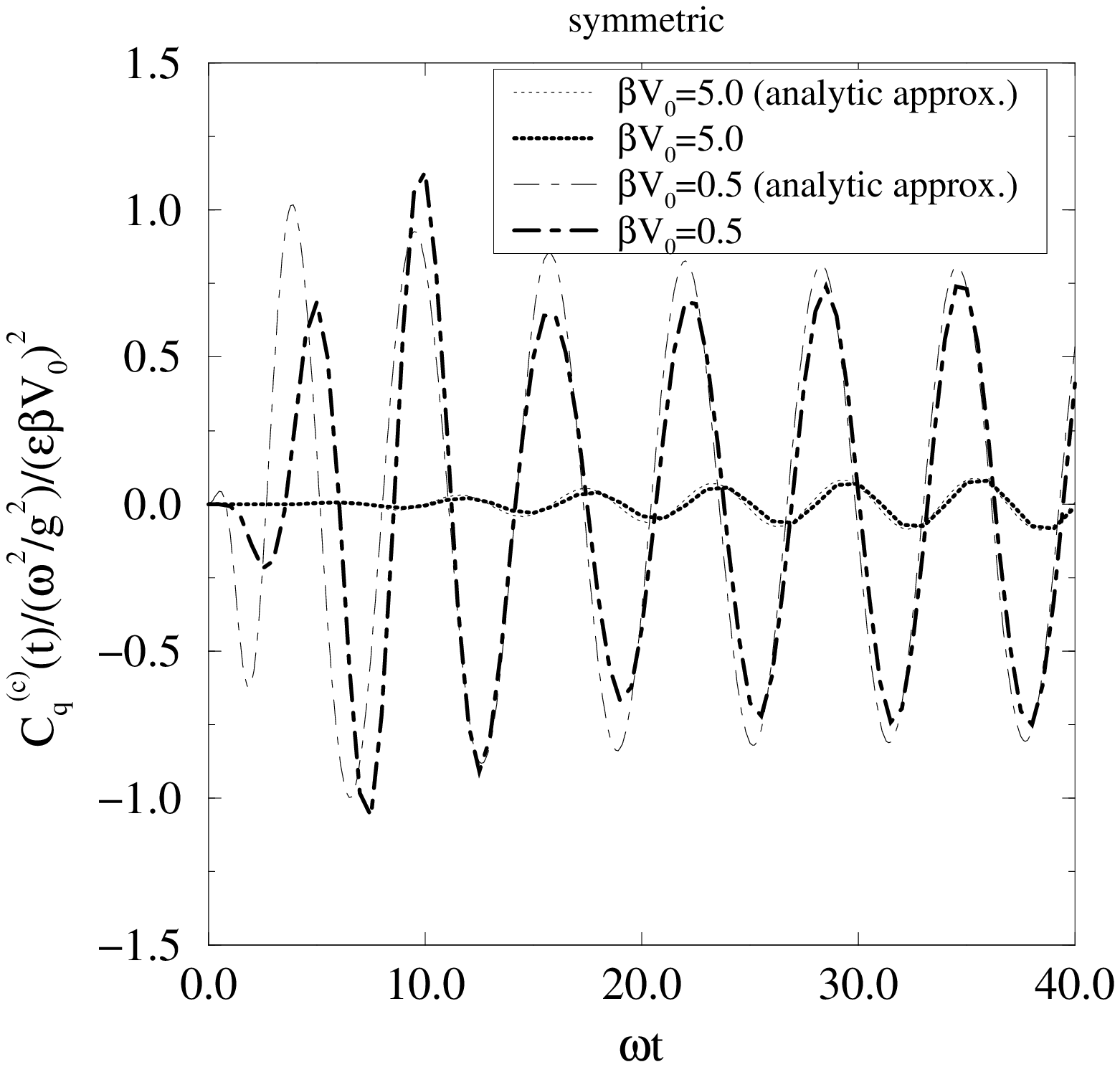}}
 
\vspace*{-6cm}
 
\caption[a]{
The quantum correction 
$C_{\hbar^2}^{(c)}(t)$ in the symmetric case, 
compared with the analytic approximation of
Sec.~\ref{ltl}.}
\la{schc}
\end{figure}

The qualitative features of the solution are the following: 
Both the classical solution $C_{\rm cl}(t)$  and the quantum
correction $C_{\hbar^2}^{(a)}(t)+C_{\hbar^2}^{(b)}(t)$ approach
zero at large times. The time scale it takes for the amplitude
to diminish depends on $\hb$, being roughly 
proportional to $\hb$, and being somewhat larger for the 
quantum corrections. The reason for the attenuation is the 
destructive interference of the continuum of classical 
solutions with different frequencies. 
This feature does not fully persist 
in the quantum case where the energy levels are
discrete: rather the behaviour is ``almost periodic''~\cite{dolan}
on a larger time scale. Indeed, already  
the term $C_{\hbar^2}^{(c)}(t)$ in Fig.~\ref{schc}
behaves in a manner qualitatively different 
from $C_{\hbar^2}^{(a)}(t)$, $C_{\hbar^2}^{(b)}(t)$: it has a constant
amplitude at large times. 

Let us discuss these features and their implications
in more quantitative terms.  

\subsection{The large time limit}
\la{ltl}

We are mainly interested in the 
large time behaviour of the correlation function.
Ordinary perturbation theory breaks down
for large times and is therefore excluded.
This is due to the secular terms in the perturbative series:
at lowest order the solution to the classical equations
of motion is proportional to $\cos(\omega t + \alpha)$ while the next
order contains a term proportional to $g^2 t \sin(\omega t + \alpha)$.
Thus, by dimensional analysis, 
straightforward perturbation theory only works for
\begin{eqnarray}
 \htt = \omega t \ll \frac{\omega^4}{g^2} \beta = 4\beta V_0.
\label{tPerturbative}
\end{eqnarray}

The way to avoid the 
secular terms is to use the exact
frequency $\Omega(E)$ inside the trigonometric functions appearing
in the perturbative series. The perturbative series 
for the classical solution of eq.~\nr{symmqct} is obtained 
from eq.~\nr{series}.
In the phase space integration
one then has to compute 
(after a change of variables according to eq.~\nr{intvars})
the dimensionless integrals
\begin{eqnarray}
J_{n m} (\beta, t) &=&
(\mbox{$\frac{g^2}{\omega^4}$})^{n + 1}\int_0^\infty dE e^{-\beta E} E^n 
 \cos(m\Omega(E) t), \nn \\
\overline{J}_{n m} (\beta, t) &=& (\mbox{$\frac{g^2}{\omega^4}$)}^{n + 1}
\int_0^\infty dE  e^{-\beta E}
E^n  \sin(m\Omega(E) t).
\label{Jn}
\end{eqnarray}
In general, this is difficult to do.
Fortunately, an exact evaluation 
is not necessary if one is interested in the
large time limit  $\omega t\gg 1$. Then it is sufficient to keep
only the first two terms of the low energy expansion
of the exact frequency, 
\be
\Omega(E) = \omega \Bigl[ 1 + \fr14 c_1 \hE 
+ {\cal O} (\hE^2 ) \Bigr],
\label{OmegaExpansion}
\ee
where $\hE=E/V_0$ and $c_1 = 3/4$. 
In this approximation we find
\begin{eqnarray}
  J_{n 1} (\beta, t) &\approx& n! 
  \frac{\cos(\omega t + (n + 1)\varphi)}
  {\Big((\frac{\omega^4}{g^2}\beta )^2 +  
  (c_1 \omega t)^2\Big)^{(n+1)/2}},\nn \\
  \overline{J}_{n 1} (\beta, t) &\approx &  n! 
  \frac{\sin(\omega t + (n + 1)\varphi)}
  {\Big((\frac{\omega^4}{g^2}\beta )^2 +  (c_1 \omega t)^2\Big)^{(n+1)/2}},
\label{JnExpansion}
\end{eqnarray}
where
\begin{eqnarray}
  \varphi = \arcsin\left(
\frac{c_1 \omega t}{\sqrt{(\frac{\omega^4}{g^2}\beta)^2 + (c_1\omega t)^2}}
\right).
\end{eqnarray}
{}From these expressions one can see that 
in the region $\omega t\gg \frac{\omega^4}{g^2}\beta$
the terms which have been neglected in eq.\ (\ref{JnExpansion})
are suppressed 
by at least one power of $1/(\omega t)$.
The reason is that 
each power of $E$ in the phase space integrand of 
eq.\ (\ref{Jn}) gives one
power of  $1/(\omega t)$. In the region 
$1\ll\omega t \ll \beta V_0$, the terms neglected 
are suppressed by $g^2\beta/\omega^4$, corresponding
to higher order perturbative corrections.
Note that the approximation
(\ref{JnExpansion}) is valid also for $\omega t \gg \beta V_0 $
where the perturbative expansion breaks down.
Thus the large time expansion for $C(t)$ can be obtained from
the low energy expansion of the phase space integrand.

Using this expansion, we find
for $C_{\rm cl}(t)$ for $\omega t \gg 1$, 
\begin{eqnarray}
  C_{\rm cl}(t) \approx 
\frac{1}{\hbar Z_{\rm cl}}
\frac{\omega^5}{g^4} J_{11}(\beta, t).
\label{CclAnalytic}
\end{eqnarray}
If $\hb=\frac{\omega^4}{4g^2} \beta \gg 1$, there is an overlap of the
``perturbative region'' of eq.~(\ref{tPerturbative}) and the large time
region:
for moderately large times $1 \ll \omega t \ll \frac{\omega^4}{g^2} \beta$ we 
recover the leading order perturbative result.
If $\omega t \gg \frac{\omega^4}{g^2}\beta $, 
in contrast, eq.\ (\ref{CclAnalytic})
simplifies to
\begin{eqnarray}
   C_{\rm cl}(t) \approx -  \frac{16}{9} \frac{1}{\hbar Z_{\rm cl}}
\frac{\omega^5}{g^4} \frac{\cos\omega t}{(\omega t)^2}.
\label{CclAsymptotic}
\end{eqnarray}
That is, for large times, the classical correlation function
oscillates with the 'tree level' frequency $\omega$ and with an
amplitude which decreases as $1/t^2$. Comparing eq. (\ref{CclAsymptotic})
with the corresponding result for the harmonic oscillator,
eq.\ (\ref{CclHO}), we see that eq.~(\ref{CclAsymptotic}) 
is non-perturbative since its functional form cannot be obtained 
by adding corrections multiplied by positive powers of $g^2$
to the harmonic oscillator result. 

Let us now 
note that if a resummation according to eq.~\nr{effw}
would take place, then the $\hbar^2$ quantum result should be
obtained by replacing $\omega^2\to\omega_{\rm eff}^2$
in eq.~\nr{CclAsymptotic}, that is 
\be
C_{\hbar^2}^{\rm resummed}(t) \approx 
\biggl[
1+b_1 \frac{g^2\hbar^2\beta}{\omega^2}\biggr] C_{\rm cl}(t)+
\fr29
\frac{1}{\hbar Z_{\rm cl}}
\frac{\hbar^2\beta\omega^3}{g^2} 
\frac{\sin\omega t}{\omega t},
\la{provh2}
\ee
where $b_1$ is some number. We show below that the true
$C_{\hbar^2}(t)$ is not of the form in eq.~\nr{provh2}.

We start with $C_{\hbar^2}^{(a)}(t)$. 
It was pointed out already in 
Sec.~\ref{Formulation} that this term is related to
the replacement $\omega^2\to\omega_{\rm eff}^2$ in the 
Hamiltonian appearing in the Boltzmann factor. To be more specific, 
the term $\omega^2$ in $U''(q)$ cancels in eq.~\nr{cha}
and in the limit $\omega t\gg1$ we find
\begin{eqnarray}
  \label{CaAsymptotic}
  C_{\hbar^2}^{(a)} (t) \approx 
\frac{1}{8} (\hbar g \beta)^2
  \langle q^2 \rangle_{\rm cl}
  C_{\rm cl}(t).
\end{eqnarray}
The contribution proportional to $\langle q^3 q_c(t)\rangle_{\rm cl}$, 
on the other hand, has one additional power of $E$ in the phase space 
integrand compared with the classical case and is
thus suppressed by a factor $1/(\omega t)$. From 
eq.~(\ref{CaAsymptotic}) it is obvious that the quantum correction 
$C_{\hbar^2}^{(a)} (t)$ shows the 
qualitative behaviour indicated in the first
term in eq.~\nr{provh2}: it is small compared with the classical result 
if $\beta\hbar\omega \ll 1$ and this holds even for arbitrarily
large times.

Next we consider the quantum corrections containing the derivatives
$\partial^2_q$, $\partial^2_p$ which we have denoted by 
$C_{\hbar^2}^{(b)} (t)$. These derivatives acting on 
the trigonometric functions in $q q_c(t)$  give extra 
factors of $t$.
When expanding the integrand in powers of energy one has
to count $t$ as $E^{-1}$.  
For $\omega t\gg 1$ we find
\begin{eqnarray}
  \label{CbAsymptotic}
  C_{\hbar^2}^{(b)} (t) \approx \frac{1}{48} \frac{1}{\hbar Z_{\rm cl}}
  \frac{\hbar^2\beta \omega^3}{g^2}
\left\{ 4 J_{01}(\beta, t) 
  - 9 \omega t \overline{J}_{11}(\beta, t) 
  - \frac94 (\omega t)^2 J_{21}(\beta, t) \right\}.
\end{eqnarray}
The individual terms in the curly brackets behave as
$\sin(\omega t)/(\omega t)$ for large times, which is the 
expected behaviour in eq.~\nr{provh2}. Such a result would
at the same time indicate that without resummation, the 
semiclassical expansion breaks down for
\begin{eqnarray}
  \omega t\gsim \frac{\omega^3}{\hbar g^2}\frac{1}{\beta\hbar\omega},
\end{eqnarray}
when the correction term in eq.~\nr{provh2}
is as large as the leading term. 
However, we find that this does not occur: in the limit
$\omega t \gg 1$ the individual terms in the curly brackets in
eq.\ (\ref{CbAsymptotic}) cancel at leading order in 
$1/(\omega t)$. Therefore the amplitude of $C_{\hbar^2}^{(b)} (t) $
decreases as $1/(\omega t)^2$ for large times. Thus
$C_{\hbar^2}^{(b)} (t)$ is small compared with the classical result
at high temperatures. The corresponding suppression factor, however,
is not in general given  by $(\beta\hbar\omega)^2 $.
There are terms proportional to
$1/(\omega t)^2$ having different dependences on
the temperature: expanding eq.\ (\ref{CbAsymptotic}) gives terms
$\propto \beta^2$ while the subleading terms in the low energy expansion
are proportional to $\beta$. We have not calculated these terms analytically.
The numerical result for the sum of $C_{\hbar^2}^{(a)} (t)$ and
$C_{\hbar^2}^{(b)} (t)$ is shown in Fig.\ \ref{ssum}.

Finally we consider the correction $C_{\hbar^2}^{(c)} (t)$.
The result for $\omega t\gg 1$  is 
\begin{eqnarray}
   C_{\hbar^2}^{(c)} (t) \approx \frac{9}{64}  \frac{1}{\hbar Z_{\rm cl}}
  \frac{\hbar^2}{\omega^4} (\omega t)^2 \{  J_{21}(\beta, t) 
  - \frac14 \omega t \overline{J}_{11}(\beta, t)
   \},
\end{eqnarray}
which for $\omega t \gg \frac{\omega^4}{g^2}\beta $ becomes
\begin{eqnarray}
  \label{CcAsymptotic}
   C_{\hbar^2}^{(c)} (t) \approx -  \frac{1}{12} 
\frac{1}{\hbar Z_{\rm cl}}
  \frac{\hbar^2}{\omega} \cos\omega t.
\end{eqnarray}
Thus at large times $ C_{\hbar^2}^{(c)} (t)$ oscillates with the 
``tree
level frequency'' but with a time independent amplitude.
This behaviour is qualitatively different from the classical case. 
Comparing eqs.\ (\ref{CclAsymptotic}), (\ref{CcAsymptotic})
we see that $ C_{\hbar^2}^{(c)} (t)$ becomes as large as the classical
correlator for $t\sim t_*$ where
\begin{eqnarray}
 \omega  t_* = \frac{\omega^3}{\hbar g^2 }=\frac{1}{\epsilon},
 \la{t*}
\end{eqnarray}
and for $t>t_*$ the semiclassical approximation breaks down.

The correction in eq.~\nr{CcAsymptotic} is clearly not of the 
form allowed by eq.~\nr{provh2}. Since there is a term of a functional
form not allowed and the allowed $\sin \omega t/(\omega t)$-term 
does not emerge, we conclude that a resummation according to 
eq.~\nr{effw} does not take place in the large time limit.
Neither can one understand the result as a resummation 
with a correction factor different from that in eq.~\nr{effw}.
Since a resummation cannot be made, the semiclassical
expansion breaks down at the time given by eq.~\nr{t*}.

It can be checked from 
Fig.\ \ref{schc} that for $\omega t \gsim 10$ the analytic approximation 
for $C_{\hbar^2}^{(c)} (t)$
indeed gives quite an accurate estimate of the exact numerical result.

To conclude, let us point out that the qualitative features found, 
together with the ``almost periodic'' behaviour~\cite{dolan}
at time scales $t\gsim t_*$, can be reproduced with the 
following approximation. Writing the full quantum result 
in eq.~\nr{c} in the energy basis, one gets
\be
C(t)=Z^{-1} \mathop{\rm Re}
\sum_{m,n}e^{-\beta E_m}e^{\frac{i}{\hbar} t(E_m-E_n)}
|\langle m|Q|n\rangle|^2. \la{enbasis}
\ee
Approximating the energy levels to first order in $g^2$, 
\be
E_n =\hbar\omega\left(n+\fr12\right)
+\fr38\frac{g^2\hbar^2}{\omega^2}\left(n^2+n+\fr12\right),
\ee
and the eigenstates to zeroth order, one gets
\be
C(t)\simeq
Z^{-1}\frac{\hbar}{2\omega}
\sum_{m=0}^{\infty}
(m+1)
\left(e^{-\beta E_{m+1}}+e^{-\beta E_{m}}\right)
\cos\! \left(\frac{E_{m+1}-E_m}{\hbar}\right)t.
\ee
The behaviour of this solution for small $\epsilon=\hbar g^2/\omega^3$
follows 
the classical
solution in Fig.~\ref{sccl}
until the time scale is of order $t\sim 4 t_*=4/\epsilon$, 
but then the periodicity sets in so that 
at the time scale $t\sim 8t_*$, the structure around $t=0$ in the 
classical solution is repeated. This is the reason for 
the breakdown of the classical approximation.  

\section{The broken case}
\la{broken}

\subsection{Preliminaries}

In the broken case, the classical Hamiltonian is
\be
H=V_0\Bigl[2\hp^2+(\hq^2-1)^2\Bigr]. \la{bH}
\ee
There exists, of course, an enormous
literature on this system.
In the present finite temperature context, it has been 
previously studied by Dolan and Kiskis~\cite{dolan}
and by Bochkarev~\cite{bochkarev}.

Quite a lot is known about the qualitative
behaviour of $C(t)$. In general, 
the solution can be written as in eq.~\nr{enbasis}.
Since the solution is a sum 
of periodic contributions corresponding to the different
energy levels that can be excited, $C(t)$ is 
``almost periodic''~\cite{dolan}. In particular, 
the lowest frequency appearing is determined by
\be
\Delta E = E_1-E_0 
\propto \hbar\omega \exp \left(-\frac{2\sqrt{2}}{3}
\frac{\omega^3}{g^2\hbar}\right),
\la{pol2}
\ee
implying that the symmetry is restored 
already at $T=0$~\cite{polyakov} in the sense that the
correlator averaged over a long enough time period vanishes.
In contrast, the
classical result $C_{\rm cl}(t)$ has a non-zero limiting value
for $t\to \infty$, in which the symmetry is only partially
restored and all the oscillations die out~\cite{dolan}. The oscillations
die out, like in the symmetric case, due to the destructive interference of
the continuum of classical solutions with different frequencies.

The fact alone that the classical result does not show
the expected qualitative behaviour of the full result, 
indicates that the classical result is not generically applicable.   
We study this problem in more concrete terms below
by evaluating the $\hbar^2$-corrections.

Note that, in seeming contrast to what was just pointed out, 
the system in eq.~\nr{bH} has also been used
to illustrate that the classical 
approximation {\it is} applicable 
to some real time problems~\cite{rs}.
The reason for the difference is that  
the situation we consider is 
different from the one in~\cite{rs}: we have a 
strict equilibrium situation
at a finite temperature $\beta^{-1}$, which is also what is considered
in~\cite{dolan,bochkarev} and which occurs 
in the real time sphaleron rate simulations. The 
consideration in~\cite{rs}, in contrast, 
concerns a non-equilibrium symmetry-restoring rate obtained 
by taking an initial state where the system is prepared in one
of the minima. 
In the strict equilibrium case, one cannot define such a rate. 
Still, the problem of the general applicability 
of the classical approximation 
to real-time problems remains. 

\subsection{Numerical Results}

The form of the classical solution $q_c(t)$ for the broken case
is discussed in~\ref{appB}.
The numerically evaluated classical correlator $C_{\rm cl}(t)$ 
is shown in Fig.~\ref{bccl}, and the quantum correction 
$C_{\hbar^2}^{(a)}(t)+C_{\hbar^2}^{(b)}(t)$ in Fig.~\ref{bsum}.

The most notable difference with respect to the symmetric case
is that there is a constant part in the broken case results. 
The energy integrand for $C_{\rm cl}(t)$ is for illustration 
shown in Fig.~\ref{encl} where 
the emergence of the constant part from $\hE<1$ can be seen.
It is evident from Fig.~\ref{bccl} that
the partial symmetry restoration in the classical result 
is the stronger
the higher the temperature is~\cite{dolan}, and 
from Fig.~\ref{bsum} that there is a further
symmetry restoring effect from the quantum corrections.   
It is seen in Fig.~\ref{bsum} that at high temperatures
($\beta V_0\sim 0.5-2.0$) the quantum corrections are
roughly proportional to the naive expansion parameter
$(\epsilon\beta V_0)^2=(\fr14\beta\hbar\omega)^2$ 
which has been factored out.

\begin{figure}[tb]
 
\vspace*{-1.0cm}
 
\hspace{1cm}
\epsfysize=18cm
\centerline{\epsffile{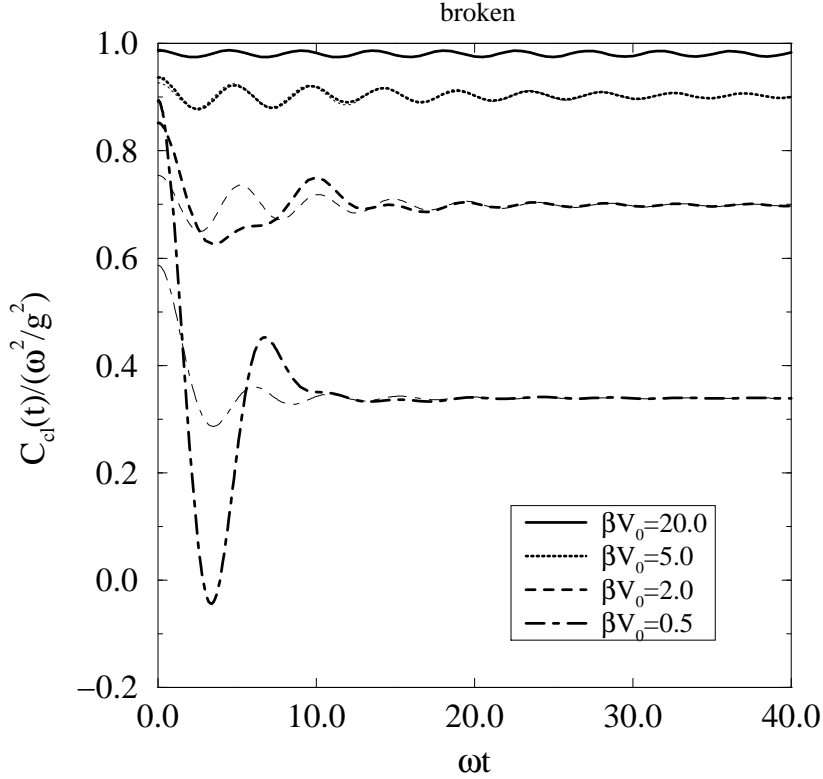}}
 
\vspace*{-6cm}
 
\caption[a]{
The classical correlator $C_{\rm cl}(t)$ in the broken 
case (thick lines), together with the analytic approximation
of Sec.~\ref{bltt} (thin lines).}
\la{bccl}
\end{figure}

\begin{figure}[tb]
 
\vspace*{-1.0cm}
 
\hspace{1cm}
\epsfysize=18cm
\centerline{\epsffile{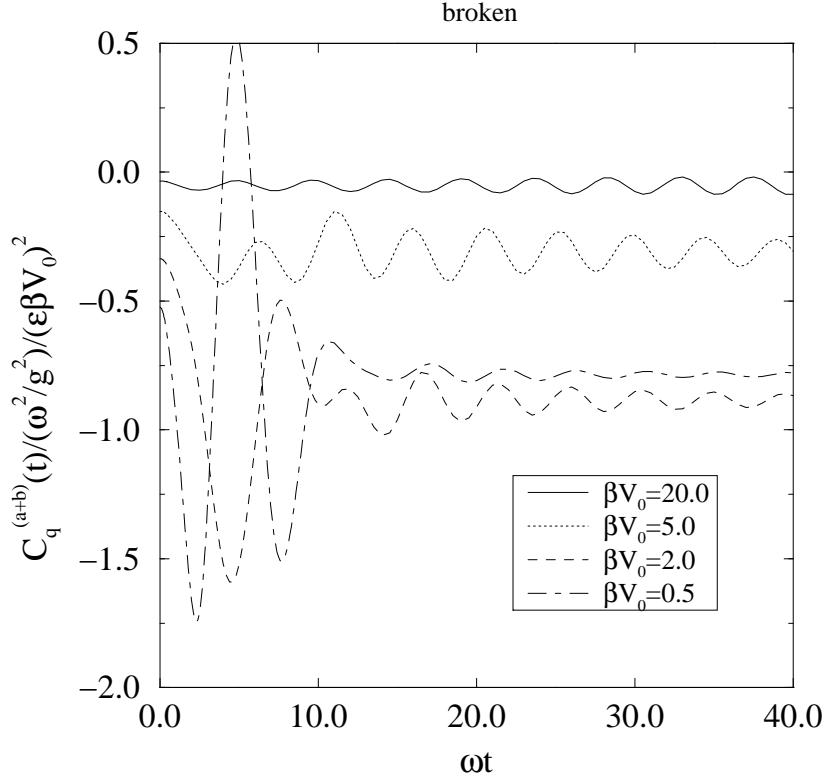}}
 
\vspace*{-6cm}
 
\caption[a]{
The quantum correction 
$C_{\hbar^2}^{(a)}(t)+C_{\hbar^2}^{(b)}(t)$ in the broken case.}
\la{bsum}
\end{figure}

\begin{figure}[tb]
 
\vspace*{-1.0cm}
 
\hspace{1cm}
\epsfysize=18cm
\centerline{\epsffile{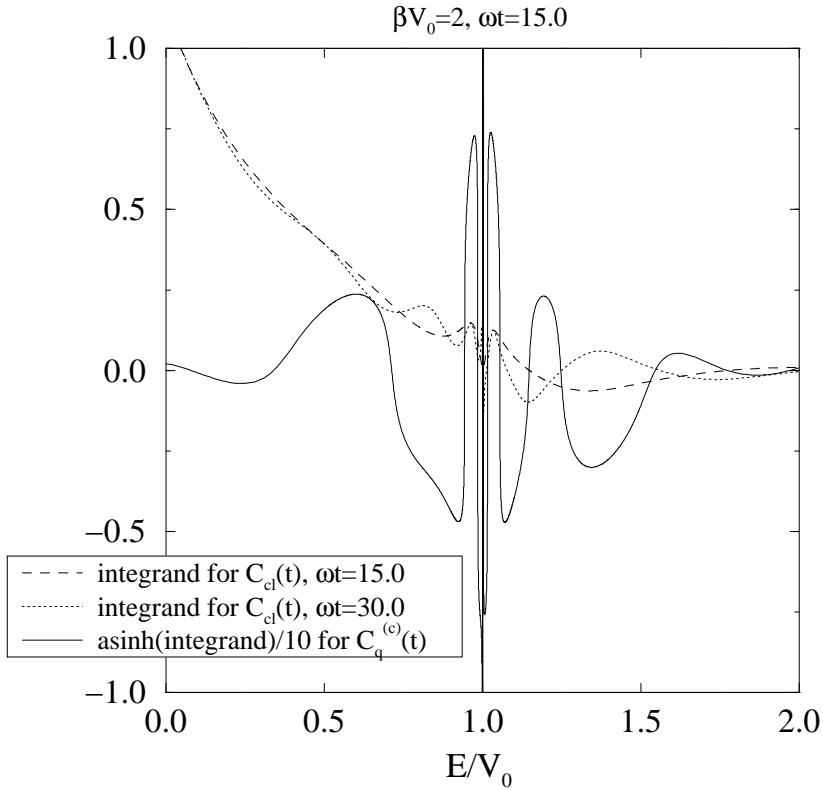}}
 
\vspace*{-6cm}
 
\caption[a]{
The energy integrands for $C_{\rm cl}(t)$ 
and $C_{\hbar^2}^{(c)}(t)$
in the broken symmetry case for $\beta V_0=2$. 
In the limit $\omega t\to\infty$ only the 
region $E/V_0<1$ contributes in $C_{\rm cl}(t)$.
The energy integrand for 
$C_{\hbar^2}^{(c)}(t)$ 
has been shown on a logarithmic scale. It
involves essentially the second derivative
of the classical integrand, which is why it has very
high peaks ($\sim 10^{10}$ already at $\omega t\sim 15$) around
$E/V_0=1$.}
\la{encl}
\end{figure}

\begin{figure}[tb]
 
 
\hspace{1cm}
\epsfysize=18cm
\centerline{\epsffile{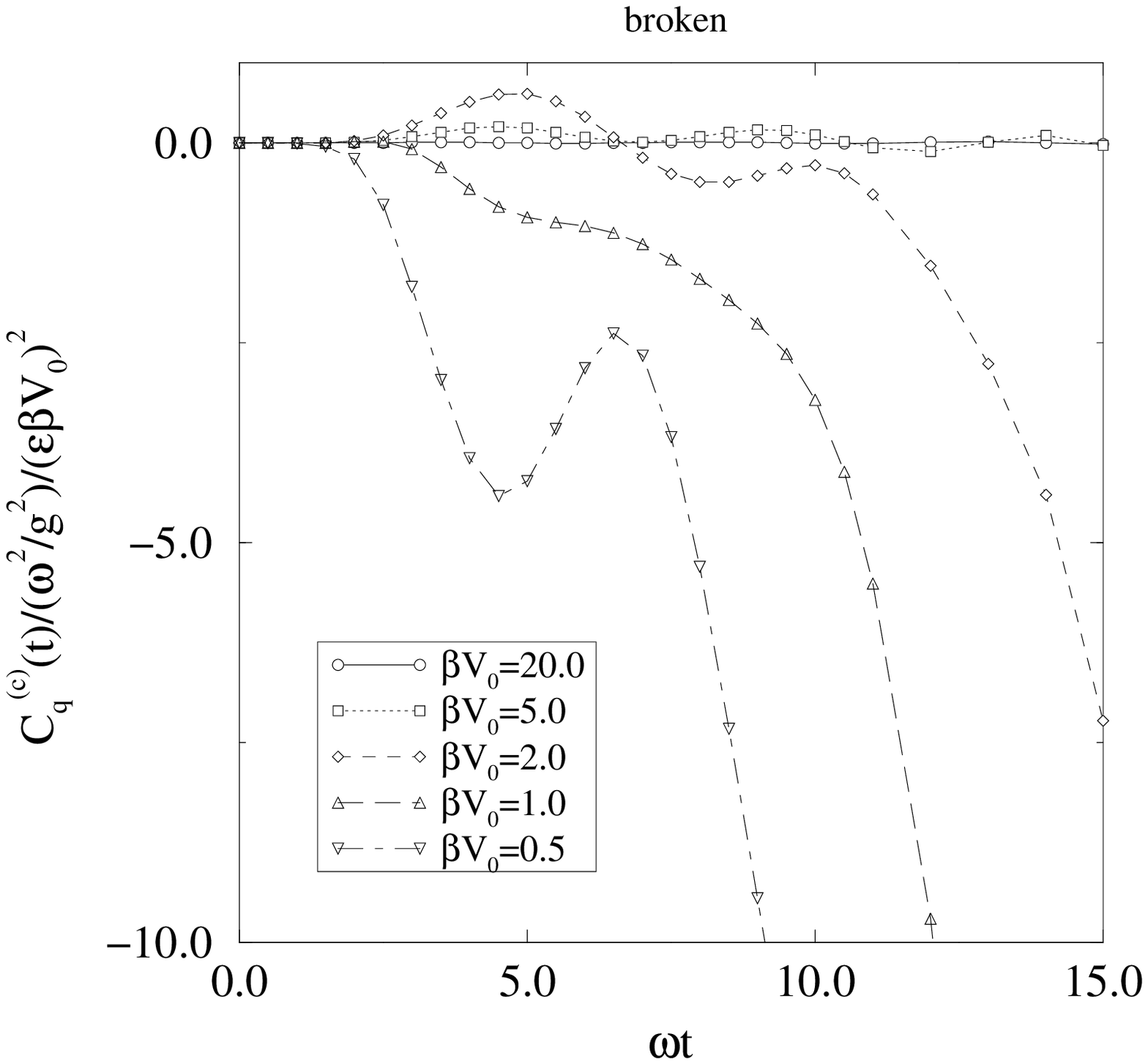}}
 
\vspace*{-6cm}
 
\caption[a]{
The quantum correction 
$C_{\hbar^2}^{(c)}(t)$ in the broken case, 
divided by $(\epsilon\beta V_0)^2=(\fr14 \beta\hbar\omega)^2$.}
\la{bchc}
\end{figure}

Let us then discuss $\hat C_{\hbar^2}^{(c)}(\hat{\beta},\htt)$.
Its numerical evaluation turns out to be very difficult 
for large $\htt$. The reason is that 
the energy-integrand is highly peaked 
and oscillatory around unity. 
To see this, note first that
at $\htt=0$, the integrand 
in eq.~\nr{sfchc} vanishes. 
Moreover, the integrand involves 
terms $\sim\sin \hat\Omega(\hE) \htt$, 
in analogy with eq.~\nr{intdndn} below.  
Hence a particular energy region will contribute provided
that 
\be
\hat\Omega(\hE) \htt \gsim 1. \la{est}
\ee 
Let $y=|\hE-1|\ll 1$. Since $K(k)\sim \ln(4/k')$ close
to $k=1$ ($k'=\sqrt{1-k^2}$), 
one gets from eq.~\nr{bfreq} that
\be
\hat\Omega(\hE)=\left\{
\begin{array}{ll}
\frac{2\pi}{\ln(64/y)}, & \hE\lsim 1 \\
\frac{\pi}{\ln(64/y)}, & \hE\gsim 1 
\end{array}
\right. . \la{omegaatone}
\ee
Eq.~\nr{est} shows then that the energy-integrand
can be large in the region 
\be
y \gsim e^{-\htt}.
\la{large}
\ee
On the other hand, the second partial derivative in eq.~\nr{sfchc} 
will involve
\be 
\partial_{\hp}^2 \Omega(\hE) = 
( \partial_{\hp} {\hE} )^2
\partial_{\hE}^2 \Omega(\hE)+\ldots,
\ee 
where $\partial_{\hp}\hE=4\hp$.
Hence according to eqs.~\nr{omegaatone}, \nr{large},
\be
\frac{\partial^2\hq_c(\htt')}{\partial \hp^2} \sim \frac{1}{y^2}\lsim
e^{2 \htt}.
\ee
Thus the height of the peaks around $\hE =1$ grows exponentially
with time, and the peaks move closer to $\hE =1$. The width 
of the peaks is diminishing, but their height is growing faster
so that they give an increasing contribution. 
In fact, the highest peak's contribution from $\hE < 1$
(where the peak gives a positive contribution) and from
$\hE>1$ (where it gives a negative one) to a large extent 
cancel, but the cancellation is not complete and one has
to account for it very precisely in the numerics to get
the remaining contribution correctly. This is 
why we cannot go to large $\htt$.
In practice, we can reliably calculate 
$\hat C_{\hbar^2}^{(c)}(\hb,\htt)$ only up to $\htt=15$, 
when the highest peaks in the energy integrand are 
of height $\sim 10^{10}$ (for $\beta V_0\sim 2$)
at $\delta\hE \sim 10^{-5}$
around unity. 
The energy integrand is shown in Fig.~\ref{encl}
and the result of the integration
in Fig.~\ref{bchc}.

\subsection{The large time limit}
\la{bltt}

Consider first the classical correlation function. 
The form of the solutions in eq.~\nr{bct} can 
be read off from eq.~\nr{series}. It is seen that 
for $\hE<1$, $q_c(t)$ contains a constant part in addition to
the cosines. 
The $\phi$-integral
obtained with the change of variables in eq.~\nr{intvars}, 
gives then 
\be
\int_{-K(k)}^{K(k)} \!d\phi\, \dn_k(\phi)\dn_k(w\htt+\phi)=
\frac{\pi^2}{2 K(k)}\biggl\{
1+8\sum_{n=1}^\infty \frac{q^{2n}}{(1+q^{2n})^2}
\cos\!\left[{n\Omega(E)t}\right]
\biggr\}.
\la{intdndn}
\ee
The cosines in eq.~\nr{intdndn} give contributions
which vanish in the limit $t \to\infty$
as in eq.~\nr{JnExpansion}, see below. Hence one gets 
from eqs.~\nr{cct}, \nr{intvars} that 
the constant part surviving is
\be
 C_{\rm cl}(t\to\infty) =\frac{\pi}{4} \frac{1}{\hbar Z_{\rm cl}}
\frac{\omega^5}{g^4} 
\int_0^1 d\hE e^{-\hb\hE}\frac{w}{K(k)}. \la{tlim}
\ee

One may also try to compute the time dependent 
part for $\omega t\gg 1$ in the same
way as in the symmetric case. There we saw that the limiting behaviour 
for large times can be obtained from a suitable low energy expansion 
of the solution to the equations of motion. In the present case it is
obvious that this expansion cannot be convergent 
when $\hE$ approaches unity. 
One may argue, however, that for large times only the solutions with small
energies are relevant and that this expansion still works. 
We find that
\begin{eqnarray}
        C_{\rm cl}(t) = C_{\rm cl}(t\to\infty) + 
        \frac{1}{16 \sqrt{2} } \frac{1}{\hbar Z_{\rm cl}}
        \frac{\omega^5}{g^4} 
        \int_0^1 d\hE e^{-\hb\hE}\Bigg( \hE 
        \cos [\Omega(\hE) t] + 
        {\cal O}(\hE^2)
        \Bigg).
\end{eqnarray} 
Replacing the upper integration limit 
by $\infty$ and keeping
only the first two terms of the low energy expansion of
\begin{eqnarray}
  \Omega(\hE) = \sqrt{2}\omega(1 - d_1 \hE +  {\cal O}(\hE^2))
\end{eqnarray}
where $d_1 = 3/16$, we obtain for $\omega t \gg 1$,
\begin{eqnarray}
        C_{\rm cl}(t) \approx C_{\rm cl}(t\to\infty) + 
        \frac{1}{16\sqrt{2} } \frac{1}{\hbar Z_{\rm cl}}
        \frac{\omega^5}{g^4} \,
        \frac{\cos(\sqrt{2}\omega t - 2 \varphi)}{(\beta V_0)^2 +
          (d_1 \sqrt{2}\omega t)^2},
\la{bAnCcl}
\end{eqnarray}
where now
\begin{eqnarray}
  \varphi = \arcsin\left(\frac{d_1 \sqrt{2}\omega t}{\sqrt{(\beta V_0)^2 +
          (d_1 \sqrt{2}\omega t)^2}}\right).
\end{eqnarray}
It can be seen in Fig.~\ref{bccl} that eq.~\nr{bAnCcl} is indeed a good
approximation at large times.

The integrals appearing in the quantum corrections
$C_{\hbar^2}^{(a)}(t)$ and $C_{\hbar^2}^{(b)}(t)$
are qualitatively quite similar to that appearing
in $C_{\rm cl}(t)$.
In particular, 
there is a constant part in these corrections
which can be evaluated in the same way
as eq.~\nr{tlim}. 
It is seen that the constant part 
tends to further restore the symmetry
compared with the classical result, see Fig.~\ref{bsum}.

For the quantum correction $C_{\hbar^2}^{(c)}(t)$, 
in contrast, the ``small energy expansion'' does not seem 
to be applicable. 
We have computed the solution but it does not agree 
with Fig.~\ref{bchc}. However, this need not be a surprise since, 
as discussed,  
it is not guaranteed that the small energy expansion
works in the broken case due to the singular nature of the 
point $\hE=1$: 
the energy integration extends beyond the radius of convergence of the 
small energy expansion. 
Moreover, the integrand in $C_{\hbar^2}^{(c)}(t)$ is qualitatively
different from that in $C_{\rm cl}(t)$.
A simple example where the small energy 
expansion would not work is given by
\be
f(\htt)=\int_0^\infty d\hE e^{-\hb\hE}
\frac{\sin \htt(1-\hE)}{\pi (1-\hE)}.
\ee
At $\htt\to\infty$ the integrand makes a delta-function, and
$f(\htt\to\infty)\to \exp(-\hb)$. Yet an expansion 
in $\hE$ of the denominator
around $\hE=0$ and an integration term by term,  
gives a result which oscillates around zero.

We could not find any other analytic way of evaluating the 
energy integral for $C_{\hbar^2}^{(c)}(t)$, 
either. The integrand is very complicated 
around $\hE\sim 1$. Thus we can only mention some 
general  features of the solution.

First, note that 
the numerical result in Fig.~\ref{bchc} shows 
that there is a growing
negative contribution at large $\htt$ in $C_{\hbar^2}^{(c)}(t)$. 
This seems to arise from $\hE$ a bit larger than unity.
To estimate very roughly when 
this kind of a contribution can be important, note
that then the peak heights must be 
such that the exponential 
suppression cannot hide them any more, that is  
\be
e^{-\hb} e^{ 2 \htt} \gsim 1.
\ee
Hence one starts to get an effect at $\htt \gsim \hb$.

As to the functional form of the solution, 
it looks roughly like $-\htt^4$ at large $\htt$. 
It is easy to see that a linear in $\htt$ behaviour
cannot occur, since it follows directly 
from the definition in eq.~\nr{c} that $C(t)$
is symmetric in $t$.
For $\beta V_0=0.5,1.0$ in which case the asymptotic
behaviour is obtained 
earliest, the leading term of 
$C_{\hbar^2}^{(c)}(t)$ can be
fitted at $\omega t\sim 8 \ldots 15$ for instance as 
\be
C_{\hbar^2}^{(c)}(t)\sim \frac{\omega^2}{g^2}\epsilon^2
\Bigl[-0.01 (\omega t)^4 \Bigr].
\la{fit}
\ee

The conclusions one can draw from the broken case
seem rather similar to those from the symmetric
case. The quantum correction $C_{\hbar^2}^{(c)}(t)$
behaves in a manner qualitatively different form 
what was observed for $C_{\rm cl}(t)$. Moreover, the
difference is such that it cannot be accounted for 
by a simple resummation of the mass parameter~$\omega^2$.
As the classical result in Fig.~\ref{bccl} is of 
order unity and the fit in eq.~\nr{fit} would suggest
the behaviour $\epsilon^2 (\omega t)^4$ for the 
quantum correction, one would expect that the 
semiclassical expansion breaks down at 
\be
\omega t \sim \omega t_* = \frac{1}{\sqrt{\epsilon}} = 
\biggl(\frac{\omega^3}{g^2\hbar}\biggr)^{1/2}.
\ee
In eq.~\nr{t*} in the symmetric case it was rather
observed that $\omega t_* = 1/\epsilon$. However, 
the fit in eq.~\nr{fit} should not be taken very 
seriously as the interval is very small, and the main 
point is that the time scale for the 
breakdown seems to be determined by~$1/\epsilon$.

Finally, let us point out that
from the general form of eq.~\nr{enbasis},
one might have expected that at finite temperature
the asymptotic values of $C(t)$ are
oscillating between positive and negative values. 
At zero temperature the time scale would be 
$\sim\exp[2\sqrt{2}/(3\epsilon)]$ according to eq.~\nr{pol2}.
Thus the quantum correction $C_{\hbar^2}^{(c)}(t)$ 
seems to restore some of the qualitative features 
missing in $C_{\rm cl}(t)$, in the sense that the 
behaviour in Fig.~\ref{bchc} looks like the beginning
of an oscillation with a large time scale. The difference
from the zero temperature case,
however, is that the time scale associated with 
the oscillation is not exponential.

\section{Summary and Conclusions}
\la{concl}

We have studied the classical 
finite temperature real time two-point correlation 
function and its first quantum corrections 
for the anharmonic oscillator.
The expansion around the classical limit
is made in powers of $\hbar$, so that each order
contains all orders in the coupling constant $g^2$.

One can identify three different 
time scales in the results. In the 
symmetric case (Section \ref{symmetric}), these are 
\be
\omega t \sim 1,\quad \omega t \sim \hb\equiv \frac{\omega^4}{4g^2}
\beta,\quad \omega t \sim \omega t_* = \frac{1}{\epsilon}\equiv
\frac{\omega^3}{g^2\hbar}.  
\ee 
As long as $\omega t \ll
\hb$, perturbation theory works and the correlation function
oscillates with period $\omega t \sim 1$.  In the non-perturbative
region $\omega t \gsim \hb$, the correlation function approaches its
asymptotic form.  We have developed a large time expansion which
allows to address also the time scales $\omega t\gg \hb$.  In this
regime the amplitude of the oscillations in the classical result
attenuates due to the destructive interference of solutions to the
equations of motion with different energies. This attenuation
cannot be associated with a damping rate.  Finally, the time scale
$t_*$ is associated with the quantum corrections and becomes infinity in the
formal limit $\hbar\to 0$. There is a hierarchy $\omega t_* \gg
\hb$ provided that $\beta\hbar\omega\ll 1$.

The general result  of our study is that at the 
non-perturbative time scales $\omega t\gsim \hb$, the 
form of the quantum corrections differs qualitatively
from that of the classical result. The 
semiclassical expansion breaks down at $t \sim t_*$
when the quantum corrections become as large as the 
classical result. Moreover, we found that these large corrections 
cannot be resummed by modifying the parameters of the classical
theory.

On the other hand,
the first quantum corrections to the classical correlation 
function are small for $t \ll t_*$. From 
this we would expect that in this region the
classical limit gives a good approximation for the full
quantum mechanical correlation function. 
The expansion parameter for the quantum corrections
in this region is not
just the naive one $(\beta\hbar\omega)^2$, but 
$\epsilon (\beta\hbar\omega)$ and $\epsilon^2$ appear, as well.

An essential question is then which of the discussed
features might be carried over to field theory. 
Unfortunately, we cannot say very much about this.
However, certainly the present study does not encourage 
one to believe in the generic applicability of the 
classical approximation in the high temperature limit
for time-dependent quantities at arbitrarily large times. 
On the other hand, there are also obvious features which
cannot hold in a four-dimensional field theory: for instance, we found that
the time $t_*$ does not depend on the temperature. This
is unlikely to be true in the pure SU(2) theory, say;
dimensionally, the classical time scale not involving
$\hbar$ is $(g^2 T)^{-1}$ in that case and the time scale
proportional to $\hbar^{-1}$ is $(\hbar g^4 T)^{-1}$.

It would be interesting to extend the present 
type of an analysis to field theory to be able to make
more concrete conclusions. Unfortunately, a straightforward
evaluation of the quantum correction $C_{\hbar^2}^{(c)}(t)$
was numerically quite demanding even in the present case, 
in particular for the ``broken'' case where the modes
with $E/V_0\sim 1$ are rather singular. In the field theory
case, the partial derivatives of the classical solution 
with respect to the initial conditions would be replaced by 
functional derivatives, making things even more complicated.
Still, one might hope that the scalar
field theory analogue of the symmetric 
case would allow a non-perturbative investigation of the
quantum corrections in the damping rate.

Finally, let us point out that as it appears
that the classical approximation does not describe the
large time behaviour at least in the present case, 
it would perhaps be useful to consider the feasibility of 
other approaches. In principle
the problem can be solved non-perturbatively
using Euclidean simulations and spectral function 
techniques. The anharmonic oscillator considered
in this paper might be a suitable toy model for developing
techniques for such studies, since it appears
that there is some non-trivial structure even in this case.

\section*{Acknowledgements}

D.B is grateful to M.Shaposhnikov and
M.L to K.Kajantie for discussions. 

\appendix
\renewcommand{\thesection}{Appendix~\Alph{section}}
\renewcommand{\theequation}{\Alph{section}.\arabic{equation}}

\section{}
\la{appA}

In this appendix, we give some details 
concerning the classical solution $q_c(t)$ and
the computation of the quantum corrections to $C_{\rm cl}(t)$
in the symmetric case.
We use the rescaled variables defined in eq.~\nr{dimless}.

The classical Hamiltonian is
\be
H=V_0\Bigl[2\hp^2+\hq^4+2\hq^2 \Bigr].
\ee
Let us introduce 
some useful notation:
\ba
k^2 &=& \frac{2\sqrt{1+\hE}}{\sqrt{1+\hE}-1} \in (2,\infty), \quad
k'^2=1-k^2<0,\quad \tilde{k}=k^{-1}<1, \nn \\
\qmax^2 &=&\sqrt{1+\hE}-1,\quad
w=\frac{\qmax}{\sqrt{2}}. \la{snotation}
\ea
The solution of the classical equations of motion is then of the form
\be
\hqc =\qmax \cn_{\tilde k} (kw\htt+k\phi), \la{symmqct}
\ee
where $\cn_k(u)$ is a Jacobi elliptic function (see \ref{appC}).
Here the initial conditions of $\hqc$ at $\htt=0$ are given by
\be
\hq_c(0)=\hq=\qmax\cn_{\tilde k}(k\phi),\quad
\dot{\hq\,}_c(0)=
\hp=-\qmax w k \,\sn_{\tilde k}(k\phi) \dn_{\tilde k}(k\phi).
\ee
For $\hq>0$ these can be inverted to give 
\be
\phi = -\mathop{\rm sign}(\hp) \tilde{k} 
F\biggl(\arcsin\!\sqrt{1-\frac{\hq^2}{\qmax^2}},\tilde{k}\biggr), \la{phi}
\ee
where $F(\phi,k)$ is the normal elliptic integral
of the first kind (see \ref{appC}). If $\hq<0$, one should
replace the result in eq.~\nr{phi} by
\be
\phi \to 2\, {\rm sign} (\phi) \ReK -\phi,
\ee
where
\be
\ReK \equiv\left\{
\begin{array}{ll}
K(k), &  k<1 \\
\tilde{k}K(\tilde{k}), & k>1
\end{array}
\right. 
\ee
(the case $k<1$ is relevant in~\ref{appB}).
Note also that
the frequency of the classical solution depends now, 
in contrast to the harmonic case, on the energy:
according to \ref{appC}, 
the period of $\cn_{\tilde k}$ is $4K(\tilde k)$ so that
it follows from eq.~\nr{symmqct} that
\be
\hat\Omega(\hE)\equiv\frac{2\pi}{P(\hE)}=
\frac{\pi w}{2 \tilde{k} K(\tilde{k})},
\ee
where $P(\hE)$ is the period of $\hqc$.

Given the classical solution, the classical 
correlation function
$C_{\rm cl}(t)$  can be calculated from eq.~\nr{cct}. 
The classical
partition function of eq.~\nr{cz} is given by 
\be
\hat Z_{\rm cl}(\hb)=\sqrt{\frac{\pi}{2\hb}}
\int_0^\infty d\hq e^{-\hb(\hq^4+2\hq^2)}=
\frac{1}{4}\sqrt{\frac{\pi}{\hb}}e^{\hb/2}K_{\fr14}(\hb/2),
 \la{sfzc}
\ee
where $K_\nu(x)$ is the modified Bessel function of the second kind.
The quantum corrections in eq.~\nr{cq} are given by
\ba
\hat C_{\hbar^2}^{(a)}(\hb,\htt) &  = & 
2 \hb^2 
\int_{-\infty}^{\infty}d\hp \int_0^{\infty}d\hq
e^{-\hb\hE} \hq^2 
[{\hat Z_{\rm cl}^{-1}}{\hat C_{\rm cl}(t)}
-\hq\hq_c(\htt)], \\
\hat C_{\hbar^2}^{(b)}(\hb,\htt) & = & 
\frac{4}{3}\hb^2
\int_{-\infty}^{\infty}d\hp  \int_0^{\infty}d\hq
e^{-\hb\hE}
\Bigl[
-(3\hq^2+1) \nn \\
& & \hspace*{1.0cm} +2\hb\hq^2(\hq^2+1)^2+2\hb\hp^2(3\hq^2+1)
\Bigr]\hq\hqc. \la{sfcha}
\ea
To get eq.~\nr{sfcha} partial integrations with respect to 
$\hp$ and $\hq$ were performed.

Concerning $\hat C_{\hbar^2}^{(c)}(\hb,\htt)$, it is useful
to make a canonical transformation 
to variables at time $\htt'$
to evaluate the Poisson bracket in eq.~\nr{chc}, 
to change then the time integration
variable and to perform one partial integration with respect to $\hp$. 
The result can be written as
\be
\hat C_{\hbar^2}^{(c)}(\hb,\htt) = 
\frac{1}{4}
\int_{-\infty}^{\infty}d\hp  \int_0^{\infty}d\hq
e^{-\hb\hE}\int_0^{\htt} d{\htt}'
\Bigl[4\hb \hp \hq_c(\htt'-\htt)-
\frac{\partial\hq_c(\htt'-\htt)}{\partial \hp}
\Bigr]\hq\frac{\partial^2\hq_c(\htt')}{\partial \hp^2}. \la{sfchc}
\ee
The partial derivatives of $\hq_c(\htt')$
can be evaluated numerically, or even analytically using the 
formulas in~\cite{byrd}. As an example, the first
derivative is
\ba
\frac{\partial\hqc}{\partial\hp} & = & 
\frac{\hp}{2k^2}(2-k^2)^3 \qmax\biggl\{
-\cn_{\tilde k}v
\biggl[\frac{\qmax^2}{2}+\frac{{\tilde k}^2}{k'^2}
\sn_{\tilde k}^2v\biggr] \nn \\
& & +\sn_{\tilde k}v\dn_{\tilde k}v\biggl[
\frac{E_{\tilde k}(v)-E_{\tilde k}(k\phi)}{k'^2}+\frac{wt}{k}\biggl(
\qmax^2+1\biggr) \nn \\
& & -\frac{w}{kk'^2}\frac{\hq}{\hp}
\biggl(1-\frac{\hq^2}{\qmax^2}\biggr)
-\frac{w k\qmax^2}{2} \frac{\hq}{\hp}
\biggr] 
\biggr\}_{v=k w\htt+k\phi}, \la{sdpp}
\ea
where $E_{\tilde k}(v)$ is defined in eq.~\nr{Eku}.

Finally, note that
since the variable $\phi$ appears in a simple manner
in $\hqc$ in eq.~\nr{symmqct}, 
it is convenient to make a change of integration variables. 
We go first into the 
canonical action-angle variables $(I,\alpha)$,  and
then from these into energy $E$ and the variable $\phi$, 
using
\be
\frac{\partial I(E)}{\partial E}=\frac{1}{\Omega (E)},\quad
\frac{\partial \alpha}{\partial\phi} = \frac{\Omega(E)}{w}.
\ee
The integration measure can then be written as 
\be
\int_{-\infty}^{\infty}d\hp\int_0^{\infty} d\hq \,[\ldots]
=\int_0^\infty d\hE
\int_{-\ReK}^{\ReK}d\phi \frac{1}{4w} [\ldots].
\la{intvars}
\ee

\section{}
\la{appB}

In this appendix, we describe 
the classical solution $q_c(t)$ and formulas for
the quantum corrections to $C_{\rm cl}(t)$
in the broken case. The classical Hamiltonian is
in eq.~\nr{bH}.

In accordance with eq.~\nr{snotation},
let us introduce some notation: 
\ba
k^2 & = & \frac{2\sqrt{\hE}}{1+\sqrt{\hE}} \in (0,2), \quad
k'^2=1-k^2,\quad
\tilde{k}=k^{-1}, \nn \\
\qmax^2 & = & 1+\sqrt{\hE},\quad
w=\frac{\qmax}{\sqrt{2}}.
\ea
Then the classical solution is 
\be
\hqc = 
\left\{
\begin{array}{ll}
\pm\qmax\dn_k(w\htt+\phi), &  \hE<1 \\
\qmax\cn_{\tilde{k}}(kw\htt+k\phi), &  \hE>1
\end{array}
\right. , \la{bct}
\ee
where the relation to $\hp,\hq$ is (for $\hq>0$)
\be
\phi=
\left\{
\begin{array}{ll}
-\mathop{\rm sign}(\hp) F\biggl(\arcsin\!\sqrt{\frac{1}{k^2}
\biggl({1-\frac{\hq^2}{\qmax^2}}\biggr)},k
\biggr), &  \hE < 1 \\
-\mathop{\rm sign}(\hp) \tilde{k} 
F\biggl(\arcsin\!\sqrt{1-\frac{\hq^2}{\qmax^2}},\tilde{k}
\biggr), &  \hE >1 
\end{array}
\right. . 
\ee
Note that $\hE<1$ corresponds to modes which do not cross
the barrier, whereas the $\hE>1$ modes do cross it.
The frequency of the classical solution again depends on energy
and is given by
\be
\hat\Omega(\hE)=\frac{2\pi}{P(\hE)}=
\left\{
\begin{array}{ll}
\frac{\pi w}{K(k)}, &  \hE<1 \\
\frac{\pi w}{2 \tilde{k} K(\tilde{k})}, &  \hE>1
\end{array}
\right. .
\la{bfreq}
\ee

The classical partition function of eq.~\nr{cz} is
\ba
\hat Z_{\rm cl}(\hb) = \sqrt{\frac{\pi}{2\hb}}
\int_0^\infty d\hq e^{-\hb(\hq^2-1)^2} =
\frac{1}{4}\sqrt{\frac{\pi}{\hb}}e^{-\hb/2}
\Big[\sqrt{2}\pi I_{\fr14}(\hb/2)+K_{\fr14}(\hb/2)
\Big],
\ea
where $I_\nu,K_\nu$ are modified Bessel functions.
The quantum corrections to the classical result of eq.~\nr{cct}
are
\ba
\hat C_{\hbar^2}^{(a)}(\hb,\htt) & = &  
2\hb^2   
\int_{-\infty}^{\infty}d\hp\int_0^{\infty}d\hq
e^{-\hb\hE} \hq^2 
[{\hat Z_{\rm cl}}^{-1}{\hat C_{\rm cl}(t)}-\hq\hqc], \\
\hat C_{\hbar^2}^{(b)}(\hb,\htt)  & = &  
\frac{4}{3}\hb^2
\int_{-\infty}^{\infty}d\hp\int_0^{\infty}d\hq
e^{-\hb\hE}
\Bigl[
(1-3\hq^2) \nn \\
& & \hspace*{1.0cm} +2\hb\hq^2(\hq^2-1)^2+2\hb\hp^2(3\hq^2-1)
\Bigr]\hq\hqc. 
\ea
The correction $\hat C_{\hbar^2}^{(c)}(\hb,\htt)$ is given 
by eq.~\nr{sfchc}. The partial derivatives of $\hq_c(\htt')$
with respect to $\hp$ can again be evaluated analytically. 
For example, for $\hE<1$ we get 
\ba
\frac{\partial\hqc}{\partial\hp} & = & 
\frac{\hp}{2k^2}(2-k^2)^3 \qmax\biggl\{
\dn_ku\biggl[\frac{\qmax^2}{2}-\frac{\sn_k^2u}{k'^2}\biggr] \nn \\
& & +\sn_ku\cn_ku\biggl[
\frac{E_k(u)-E_k(\phi)}{k'^2}-wt\biggl(
\frac{\qmax^2k^2}{2}+1\biggr) \nn \\
& & -\frac{1}{k'^2w}\frac{\hp}{\hq}+
\qmax^2 w\frac{\hq^2-1}{\hq\hp}
\biggr] 
\biggr\}_{u=w\htt+\phi},
\ea
and for $\hE>1$ the expression is the one in eq.~\nr{sdpp}
with the replacement $\qmax^2\to-\qmax^2$, except for the $\qmax^2$
appearing in $\hq^2/\qmax^2$.

Finally, a change of integration variables can be made 
according to eq.~\nr{intvars}.

\section{}
\la{appC}

We discuss here briefly 
some of the basic definitions of the Jacobi elliptic
functions used. The notation follows~\cite{ww,byrd}.

Let $F(\phi,k)$ be the normal elliptic integral of the first kind, 
\be
F(\phi,k)=\int_0^\phi\frac{d\alpha}{\sqrt{1-k^2\sin^2\!\alpha}}.
\ee
Then the complete
elliptic integral of the first kind $K(k)$ is defined by 
\be
K(k)=F(\frac{\pi}{2},k)=\int_0^1\frac{dt}{\sqrt{(1-t^2)(1-k^2t^2)}}.
\ee
The associated {\it nome} is 
\be
q =\exp\biggl[-\pi\frac{K(k')}{K(k)}\biggr], \la{nome}
\ee
where $k'^2=1-k^2$.
Similarly,
\ba
E(\phi,k) & = & \int_0^\phi \! d\alpha\sqrt{1-k^2\sin^2\alpha},\quad
u=\int_0^{\mathop{\rm am}_k u}\frac{d\alpha}{\sqrt{1-k^2\sin^2\alpha}},
\nn \\
E_k(u) & \equiv &  E({\rm am}_k u,k)
\la{Eku}
\ea
define the normal elliptic integral of 
the second kind $E(\phi,k)$, the amplitude function
${\rm am}_k u$, and the function $E_k(u)$. 

The Jacobi elliptic functions are denoted by
$\dn_ku,\sn_ku,\cn_ku$. 
They are defined by ($0<u<K(k)$)
\ba
u &=& \int_{\dn_ku}^1
\frac{dt}{\sqrt{(1-t^2)(t^2-k'^2)}}=
\int_{\cn_ku}^1
\frac{dt}{\sqrt{(1-t^2)(k'^2+k^2t^2)}} \nn \\
& = & 
\int_0^{\sn_ku}
\frac{dt}{\sqrt{(1-t^2)(1-k^2t^2)}}.
\ea
Note that $\sn_ku=\sin ({\rm am}_ku)$.
The periodicity of $\dn_ku$ is $2 K(k)$, and that
of $\sn_ku,\cn_ku$ is $4K(k)$; $\dn_ku, \cn_ku$
are symmetric while $\sn_ku$ is anti-symmetric.
The functions appearing in the classical solution $q_c(t)$
for the anharmonic oscillator are $\dn_ku$ and $\cn_ku$. 
We need their Fourier expansions,
\ba
\dn_ku & = & \frac{\pi}{2K(k)}
\biggl[
1+4\sum_{n=1}^\infty\frac{q^n}{1+q^{2n}}\cos\frac{n\pi u}{K(k)}
\biggr], \nonumber \\
\cn_ku & = & \frac{2 \pi}{kK(k)}
\sum_{n=1}^\infty\frac{q^{n-1/2}}{1+q^{2n-1}}
\cos\!\frac{(n-{1\over2})\pi u}{K(k)},
\la{series}
\ea
where $q$ is the nome in eq.~\nr{nome}.

\end{document}